\documentclass{iau}
\usepackage{graphicx}
\usepackage{amsmath}
\usepackage{amssymb}
\usepackage{natbib}

\newcommand{\swift}{\textit{Swift}}

\newcommand{\lum}{\mathrm{erg~s}^{-1}}
\newcommand{\flux}{\mathrm{erg~cm}^{-2}~\mathrm{s}^{-1}}
\newcommand{\fluence}{\mathrm{erg~cm}^{-2}}

\newcommand{\nh}{\mathrm{cm}^{-2}}

\newcommand{\grs}{GRS 1741--2853}

\newcommand{\ax}{AX J1745.6--2901}
\newcommand{\sgra}{Sgr A$^*$}

\def \apj {\textit{ApJ}}

\def \aap {\textit{A\&A}}

\title[IAUS303.~~\ax\ and \grs] %% short title %%
{The Galactic center X-ray transients \\ \ax\ and \grs} %% full title %%

\author[N. Degenaar et al.]  %% short author list %%
{N. Degenaar$^{1}$, R. Wijnands$^2$, M. T. Reynolds$^{1}$, J. M. Miller$^{1}$,\\ J. Kennea$^{3}$ and N. Gehrels$^{4}$ on behalf of a larger collaboration}

\affiliation{$^1$Hubble Fellow; Department of Astronomy, U. of Michigan, Ann Arbor, MI 48109, USA \\
$^2$Astronomical Institute ``Anton Pannekoek", U. of Amsterdam, 1090 GE Amsterdam, NL\\
$^3$Department of Astronomy and Astrophysics, PSU, University Park, PA 16802, USA\\
$^4$Astrophysics Science Division, NASA Goddard Space Flight Center, Greenbelt, MD, USA}

\pubyear{2014}
\volume{303}  %% insert here IAU Symposium No.
\jname{\mbox{The Galactic Center: Feeding and Feedback in a Normal Galactic Nucleus}}
\editors{Editors} 
\begin{document}

\maketitle

%% -- Abstract ----------------------------------
\begin{abstract}
\ax\ and \grs\ are two transient neutron star low-mass X-ray binaries that are located within $\simeq$10$'$ from the Galactic center. Multi-year monitoring observations with the \swift/XRT has exposed several accretion outbursts from these objects. We report on their updated X-ray light curves and renewed activity that occurred in 2010--2013.
\end{abstract}

\firstsection 
\section{Introduction}
\begin{figure}[tb]
\begin{center}
 \includegraphics[width=0.49\textwidth]{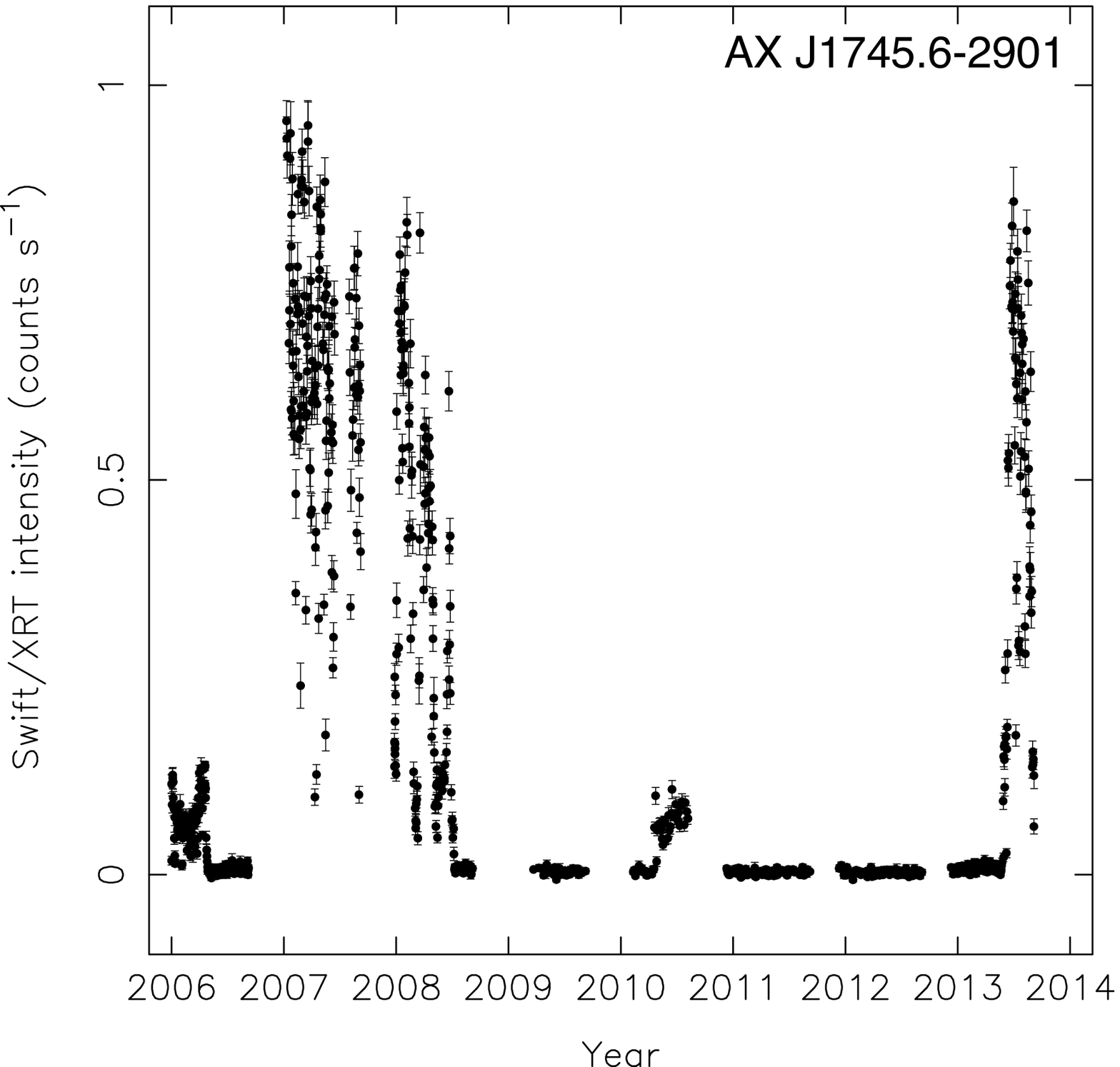} 
  \includegraphics[width=0.49\textwidth]{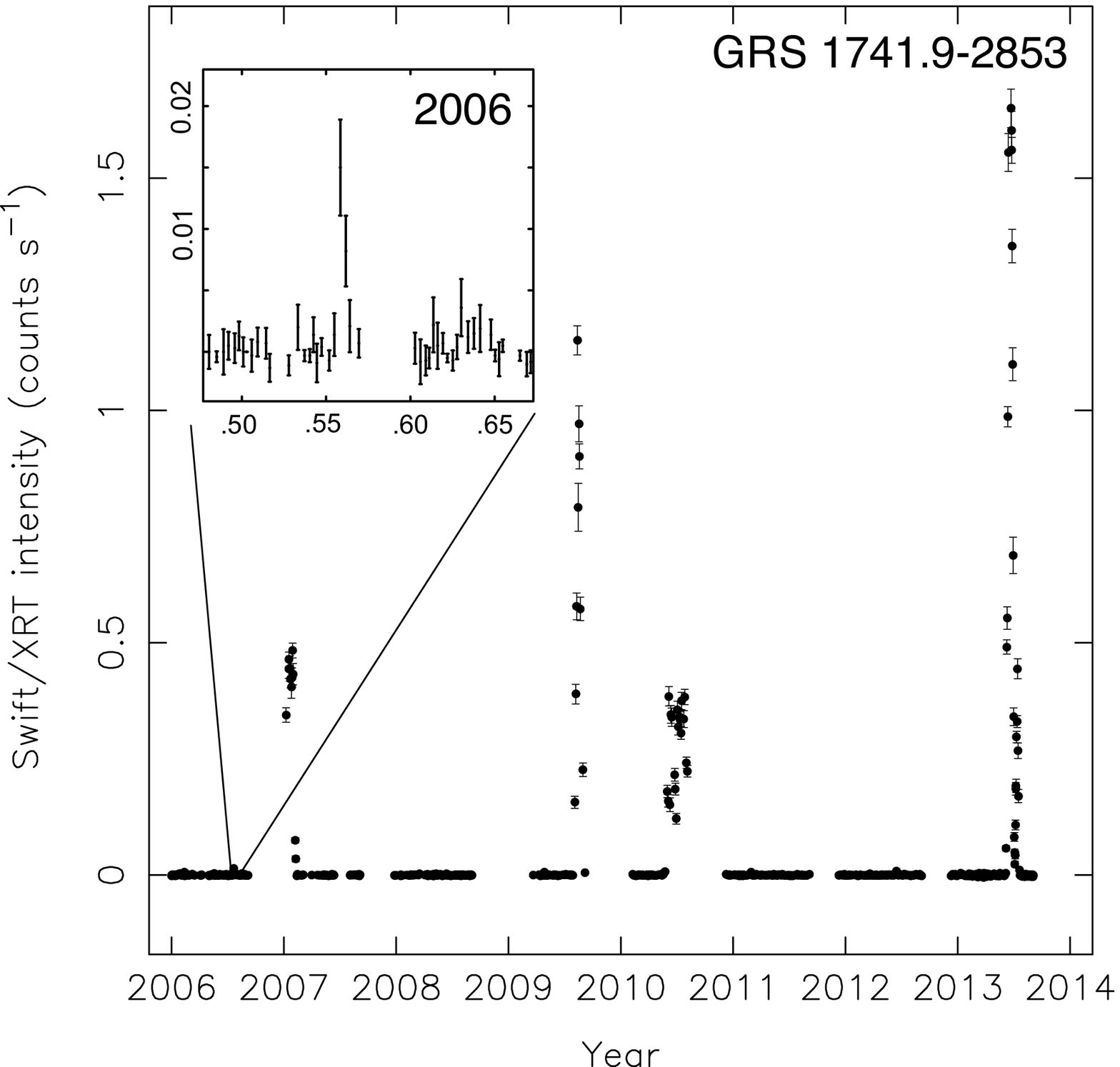} 
 \caption{{
\swift/XRT light curves of \ax\ and \grs\ (2006--2013). Both sources exhibited four main outbursts. \grs\ also displayed a very weak ($L_{\mathrm{X}}^{\mathrm{peak}} \simeq 7 \times10^{34}~\lum$) and short ($\simeq$1 week) outburst in 2006 \citep[][]{degenaar09_gc}. 
 }}
 \label{fig:longlc}
\end{center}
\end{figure}

\swift\ has monitored the inner $\simeq25'\times25'$ of the Milky Way with the onboard X-Ray Telescope since 2006, using $\simeq$1~ks exposures performed every 1--4~days. This has amounted to nearly 1000 observations and $\simeq$1.1~Ms of exposure time between 2006 and 2013. This campaign provides a perfect setup to study the long-term X-ray behavior of the supermassive black hole \sgra, as well as 15 known transient low-mass X-ray binaries \citep[e.g.,][]{degenaar09_gc,degenaar2010_gc,degenaar2013_sgra}. \ax\ and \grs\ are two neutron star low-mass X-ray binaries that are frequently active; both exhibited two main outbursts in the period 2006--2009 \citep[][see also Figure~\ref{fig:longlc}]{degenaar09_gc,degenaar2010_gc}.

\section{New Outbursts from \ax\ and \grs}\label{sec:ob}
Figure~\ref{fig:longlc} displays the updated X-ray light curves of \ax\ and \grs, including new \swift/XRT data obtained in 2010--2013. Both sources exhibited two new accretion outbursts. We fitted the average spectra of each outburst to an absorbed power-law model. We calculated the 2--10 keV fluxes and fluences, and determined the outburst peak intensity. The results are summarized in Table~\ref{tab:spec}. No X-ray bursts were detected from the two sources in the 2010--2013 \swift\ data.

\ax\ entered an outburst between 2010 June 11 and 15 and remained active till October 31, after which the Galactic center became Sun constrained. When the observations resumed on 2011 February 4, the source had faded to the background level, implying an outburst duration of $\simeq$20--34 weeks. Renewed activity was detected on 2013 July 18 and continued until the observations ended on 2013 November 5, implying an outburst duration of $\gtrsim$16~weeks. 
\grs\ was seen in outburst starting on 2010 July 18 and remained active for $\simeq$13 weeks. The source was again active for $\simeq$6 weeks from 2013 August 1 till September 14.

\begin{table}[tb]
\caption{Spectral Properties of the New Outbursts of \ax\ and \grs.}
\begin{center}
\begin{tabular}{p{2.3pc} p{3.6pc} p{3.2pc} p{3.2pc} p{3.5pc} p{3.5pc} p{2.5pc} p{2.2pc} p{3.5pc}}
\hline
Year & $N_H$ & $\Gamma$ & $\chi^2$/dof& $F_{\mathrm{X}}^{\mathrm{unabs}}$ & $L_{\mathrm{X}}$ & $L_{\mathrm{X}}^{\mathrm{peak}}$& $t_{\mathrm{ob}}$ & $f$ \\
\hline
\multicolumn{8}{l}{\textbf{\ax}}\\
2010 & $2.0 \pm 0.2$ & $2.0 \pm 0.3$ & $111/109$ & $0.4 \pm 0.1$ & $2.7 \pm 0.5$ & $\simeq5.6$ & 20--34 & $\simeq4.3-7.4$ \\
2013 & $2.0 \pm 0.1$ & $2.6 \pm 0.1$ & $567/472$ & $3.3 \pm 0.2$ & $25 \pm 1.0$ & $\simeq67$ & $\gtrsim16$ & $\gtrsim32$ \\
\multicolumn{8}{l}{\textbf{\grs}}\\
2010 & $1.3 \pm 0.8$ & $2.2 \pm 0.2$ & $252/222$ & $1.2 \pm 0.1$ & $6.3 \pm 0.4$ & $\simeq14$ & $\simeq13$ & $\simeq9.5$ \\
2013 & $1.4 \pm 0.1$ & $2.9 \pm 0.1$ & $443/434$ & $8.2 \pm 0.4$ & $44 \pm 2.0$ & $\simeq230$ & $\simeq6$ & $\simeq30$ \\
\hline
\end{tabular}
\label{tab:spec}
\end{center}
Note.-- Quoted errors refer to $90\%$ confidence levels. The hydrogen column density ($N_H$) is in units of $10^{23}~\nh$. Unabsorbed fluxes ($F_{\mathrm{X}}^{\mathrm{unabs}}$) are in units of $10^{-10}~\flux$ and for the 2--10 keV energy range. Fluences ($f$, in units of $10^{-4}~\fluence$) were estimated by multiplying the average 2--10 keV flux with the outburst duration ($t_{\mathrm{ob}}$, given in weeks). Average ($L_{\mathrm{X}}$) and peak ($L_{\mathrm{X}}^{\mathrm{peak}}$) luminosities are in units of $10^{35}~\lum$ and were calculated by assuming a distance of $D=8$~kpc for \ax\ and $D=6.7$~kpc for \grs. 
\end{table}

\section{Summary and Outlook}\label{sec:conclude}
In the past seven years, \ax\ and \grs\ have each displayed four distinct outbursts captured by the \swift/XRT (Figure~\ref{fig:longlc}). The main outbursts of \grs\ (i.e., neglecting the mini-outburst observed in 2006; Figure~\ref{fig:longlc}) have varying peak intensities and lengths, yet comparable fluencies \citep[Table~\ref{tab:spec}, see also][]{degenaar2010_gc}. \ax\ displays two types of outbursts: those in 2006 and 2010 were relatively faint ($L_{\mathrm{X}}\simeq4\times10^{35}~\lum$) and short (months), whereas the 2007--2008 outburst was much brighter ($L_{\mathrm{X}}\simeq 2\times10^{36}~\lum$) and longer (1.5 yr). The peak luminosity detected in 2013 was similar to that of the 2007--2008 outburst. If the duration is also similar, the source might still be active when the Galactic center becomes observable again in 2014 February. However, it was argued by \citet{degenaar2010_gc} that long and bright outbursts can only recur on a time scale of a decade.

\end{document}